\documentclass[aps,twocolumn,prl,showpacs,preprintnumbers,amsmath,amssymb,floatfix]{revtex4}
\usepackage{graphicx}% Include figure files
\usepackage{dcolumn}% Align table columns on decimal point
\usepackage{bm}% bold math
\usepackage[english]{babel}
\begin{document}
\preprint{APS/123-QED}
\author{T. Westerkamp$^{(1)}$, M. Deppe$^{(1)}$, R. K\"uchler$^{(1)}$, M. Brando$^{(1)}$,\\
C. Geibel$^{(1)}$, P. Gegenwart$^{(2)}$, A. P. Pikul$^{(3)}$, and F. Steglich$^{(1)}$}
\affiliation{$^{(1)}$ Max Planck Institute for Chemical Physics of Solids, 01187 Dresden, Germany\\
$^{(2)}$ I. Physikalisches Institut, Georg-August-Universit\"at, 37077 G\"ottingen, Germany\\
$^{(3)}$ Institute of Low Temperature and Structure Research,
Pol. Acad. Sci., 50-950 Wroc\l aw 2, Poland}
\title{Kondo-Cluster-Glass State near a Ferromagnetic Quantum Phase Transition}
\date{\today}
\begin{abstract}
We report on a comprehensive study of CePd$_{1-x}$Rh$_x$ $(0.6 \leq
x \leq 0.95)$ poly- and single crystals close to the ferromagnetic
instability by means of low-temperature ac susceptibility,
magnetization and volume thermal expansion. The signature of
ferromagnetism in this heavy-fermion system can be traced  from
6.6~K in CePd down to 25~mK for $x=0.87$. Despite pronounced
non-Fermi-liquid (NFL) effects in both, specific heat and thermal
expansion, the Gr\"uneisen ratio {\it does not} diverge as
$T\rightarrow 0$, providing evidence for the absence of a quantum
critical point. Instead, a peculiar "Kondo-cluster-glass" state is
found for $x\geq 0.65$, and the NFL effects in the specific heat, ac
susceptibility and magnetization are compatible with the quantum
Griffiths phase scenario.
\end{abstract}
\pacs{71.27.+a} \maketitle
%
%The physics of strongly correlated electron systems reveals a
%variety of outstanding phenomena. Particular attention is
%currently given to quantum critical points (QCPs),
%zero-temperature second-order phase transitions that are
%continuously driven by quantum instead of thermal
%fluctuations~\cite{Gegenwart08}. Compounds exhibiting a QCP are
%of strongest interest, because the presence of a QCP leads to
%non-Fermi-liquid (NFL) behavior, sometimes to unconventional
%superconductivity and perhaps to more unusual states of
%matter~\cite{Loehneysen07}.
%
The ground state of $f$-electron-based Kondo-lattice (KL) systems
depends
%(paramagnetic or magnetically ordered)
sensitively on the balance between Kondo- and exchange interactions.
While recently numerous antiferromagnetic (AF) KL systems have been
tuned towards a quantum critical point (QCP) by variation of
pressure or doping~\cite{Loehneysen07}, appropriate KL candidates
for the study of ferromagnetic (FM) QCPs are extremely
rare~\cite{Stewart}. Starting deep in the localized moment regime,
in several Ce-based ferromagnets the increase of the Kondo
interaction with pressure tends to stabilize an AF ground state
before the QCP is reached \cite{Sullow,Eichler}. Binary CePt may be
an exception \cite{Larrea}, although the FM signature is
dramatically weakened under pressure well before the ordering
temperature vanishes, and transport experiments suggest a sudden
drop of the phase boundary close to the critical pressure. This
behavior resembles the case of pure FM transition-metal compounds,
which display first-order quantum phase transitions (QPTs) under
pressure \cite{Pfleiderer97, Saxena00, Uhlarz04}.

Theoretical studies have suggested that the suppression of itinerant
ferromagnetism in clean systems, in contrast to antiferromagnetism,
always ends at a classical critical point (at finite $T$) where a
first-order phase transition occurs~\cite{Kirkpatrick03,Chubukov04}.
For KL systems it is questionable whether all QCPs could be
described in an itinerant scenario \cite{Gegenwart08}. Thus, a
detailed investigation of suitable FM systems close to their
instability is highly desired. Furthermore, theoretical calculations
show that disorder in a system may smear out the QPT, resulting in
an exponential suppression of the ordered state~\cite{Vojta03}. In
fact, experiments on doped FM materials as, e.g., the itinerant
Zr$_{1-x}$Nb$_x$Zn$_2$~\cite{Sokolov06} or the $5f$-based
heavy-fermion (HF) system URh$_{1-x}$Ru$_x$Ge~\cite{Huy07} exhibited
a continuous depression of the ordered state.

In this Letter, we investigate the KL system CePd$_{1-x}$Rh$_x$,
which is an ideal candidate to explore a FM QPT
as it provides the opportunity to investigate
the evolution of FM ordering by tuning the substitution of the Ce
ligands. The system evolves from a FM ground state in CePd with
$T_C=6.6$~K to a non-magnetic intermediate-valence (IV) state in
CeRh. The whole series crystallizes in the orthorhombic CrB
structure. The observed decrease of $T_C$ over more than two
decades in temperature, from 6.6~K at $x=0$ to 25~mK at $x=0.87$,
is presently the best known example for the continuous
disappearance of FM order in any KL
system~\cite{Kappler91,Sereni07}.

Evidence for the FM nature of the ordered state stems from the
temperature dependence of the ac susceptibility which shows sharp
maxima for all investigated samples. The competition between FM
order and, with increasing Rh content, growing Kondo screening leads
to a continuous decrease of $T_C$. Furthermore, the smaller Rh gives
rise to a volume compression of the compound's unit cell and changes
its electronic structure. Most interestingly, the curvature of the
phase boundary $T_C(x)$ changes from negative for $x<0.6$ to
positive for $x\geq 0.6$, displaying a long tail towards higher Rh
contents. In this concentration range, the Kondo temperature
$T_K\approx\Theta_p/2$ ($\Theta_p$: paramagnetic Weiss temperature)
strongly increases with $x$~\cite{Sereni07}.

Specific-heat measurements have proved the existence of NFL
behavior for concentrations close to the disappearance of FM
order~\cite{Pikul06}. At $x=0.85$ a logarithmic increase of the
specific-heat coefficient $\gamma=\Delta C(T)/T$ down to the
lowest temperature of 80~mK was observed.
%, $\Delta C(T)$ denoting the 4$f$ increment to the specific heat.
Samples with higher Rh content showed a power-law $T$-dependence,
$\gamma\propto T^{\lambda -1}$ with $\lambda=0.6$ and $0.67$ for
$x=0.87$ and $x=0.9$, respectively. For $x=0.8$, the magnetic
entropy increment $\Delta S$ is less than 0.4~$R$ln2 up to 6~K. With
increasing $x$ this value becomes drastically reduced.  An analysis
of the entropy and the temperature dependence of the susceptibility
at 2\,K revealed some fraction of still unscreened magnetic moments,
even at high $x$ where the {\it average} $T_K$ is already above
50~K. Thus, a broad distribution of local $T_K$ values with a tail
down to $T_K\rightarrow 0$ is realized in this
system~\cite{Pikul06}.

The $T$-$x$ phase diagram and the evolution of $T_K$ in
CePd$_{1-x}$Rh$_x$ raise questions concerning the mechanism
behind the suppression of FM order and the presence of a QCP
at the FM quantum phase transition. Below, we present results of
thermal expansion, ac susceptibility and magnetization
measurements in the region of the phase diagram where
ferromagnetism disappears. The experiments were performed on
poly- and single crystals that have been characterized
before~\cite{Sereni07,Deppe06}. %Our results are incompatible with
%the scaling predictions for a QCP and provide evidence
%for the formation of frozen clusters in the region where $T_C(x)$
%shows a tail.

%------thermal expansion-------------------------------------------------------------
The coefficient of volume thermal expansion,
$\beta(T)=V^{-1}(dV/dT)$ ($V$: sample volume) is more singular than
the specific heat $C(T)$ when approaching a pressure-sensitive
QCP~\cite{Zhu}. Consequently, the Gr\"uneisen ratio
$\Gamma\sim\beta/C$  must diverge algebraically in the approach of
any HF QCP, as recently found for several KL systems exhibiting a
QCP of AF nature~\cite{Gegenwart08}. Fig.\,1a shows $\beta$ of
polycrystalline CePd$_{1-x}$Rh$_x$ with $0.8 \leq x\leq 0.95$
plotted as $\beta(T)/T$ vs. $\log T$. In agreement with $C(T)/T$ and
$\chi_{ac}(T)$ results the minimum at $T\approx 0.25$~K in
$\beta(T)/T$ marks the magnetic transition for $x=0.8$. Upon
increasing the Rh concentration, $\beta(T)/T$ shows no sign of phase
transitions for $x=0.87$ and $0.9$, but rather diverges on cooling
to 0.1~K. Note that for these concentrations $\beta(T)$ is negative
and that for $x=0.9$ the divergence is even larger than for
$x=0.87$, with absolute values comparable to those found in HF
metals close to a QCP~\cite{Gegenwart08}. For the $x=0.95$ sample,
$\beta(T)/T$ is always positive, as expected for paramagnetic Ce
systems, with smaller absolute values. These results are in contrast
to those of specific-heat measurements on $0.8 \leq x \leq 0.95$
which clearly show a continuous decrease of the $C(T)/T$ values with
increasing $x$, as expected when approaching the IV
regime~\cite{Pikul06}.

Analyzing the dimensionless Gr\"uneisen ratio, defined as
$\Gamma=V_m/\kappa_T\cdot\beta/C$, where $V_m$ and $\kappa_T$ denote
the molar volume and isothermal compressibility, respectively, we
find a striking deviation from the predicted stronger than
logarithmic divergence for a QCP~\cite{Zhu}: At $x=0.87$, i.e., very
close to the Rh-concentration for which the anomaly in
$\chi_{ac}(T)$ disappears, an almost similar power-law behavior has
been found for $C(T)/T$ and $\beta(T)/T$, leaving a virtually
temperature independent Gr\"uneisen ratio (see Fig.\,1b). Thus, a
QCP scenario can be discarded. Interestingly, in the paramagnetic
regime, $x\geq0.9$, $|\Gamma(T)|$ strongly increases on cooling in
an almost logarithmic fashion and seems to saturate at the lowest
temperatures.
\begin{figure}
\centerline{\includegraphics[width=\linewidth,keepaspectratio]{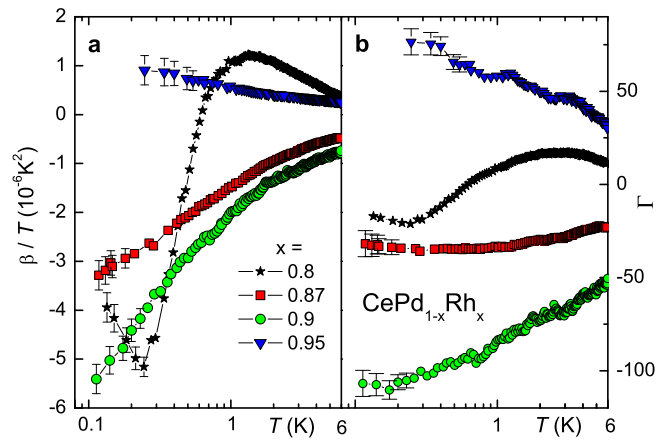}}
\caption{(color online) a: Volume thermal expansion $\beta (T)$ of
CePd$_{1-x}$Rh$_x$ polycrystals plotted as $\beta/T$ vs. $\log T$.
b: Dimensionless Gr\"uneisen ratio $\Gamma=V_m/\kappa_T\cdot\beta/C$
vs. $\log T$ with $V_m=6.6\cdot 10^{-5}$ m$^3$mol$^{-1}$ and
$\kappa_T=1\cdot 10^{-11}$ Pa$^{-1}$.} \label{fig4}
\end{figure}
The negative Gr\"uneisen ratio for $x=0.87$ and $0.9$ indicates an
unusual volume dependence. For paramagnetic Ce systems, a positive
$\Gamma$, as observed for $x=0.95$, is expected, since the Kondo
interaction, being the dominant energy scale, increases under
hydrostatic pressure. On the other hand, a negative sign is usually
associated with magnetic ordering due to the RKKY interaction, which
decreases under hydrostatic pressure. Since $\Gamma<0$ even in the
paramagnetic regime, our data suggest the presence of substantial
magnetic correlations in addition to the Kondo effect. In order to
clarify the situation, we performed detailed magnetization and ac
susceptibility measurements.
%*************ac susceptibility**************************************

Low-temperature ac susceptibility was measured down to 20~mK at
various frequencies $\nu$ on poly- and single crystals in the
concentration range $0.6\leq x\leq 0.9$. Although the absolute
values of $\chi_{ac}$ decrease with increasing Rh content, it was
possible to trace the transition temperature down to 25 mK for
$x=0.87$ (inset a of Fig.~2)~\cite{Sereni07}.
The $x=0.6$ sample clearly shows a FM phase transition at $T_C=2.4$~K.
The $x=0.9$ sample does not show any ordering down to 20~mK. The pronounced
$\chi_{ac}$ maxima of those samples with concentrations in between
exhibit a frequency dependence. Poly- and single crystals show
similar behavior. Single crystals were probed with the modulation
field $\mu_{0}h_{ac}\|c$. This transition appears to match several
indications of spin-glass-type freezing. As displayed in Fig.~2,
the $\chi_{ac}(T)$ signal of a single crystal with $x = 0.8$
shows a pronounced cusp in its real part $\chi'$ and a
corresponding inflection point in the imaginary part $\chi''$.
Both signals display a clear frequency dependence at the
temperature of the $\chi'(T)$ cusp, labeled $T_C^*$ in order to
distinguish it from the Curie temperature $T_C$ found at lower Rh
content. As in spin glasses $\chi'(T_C^*)$ is extreme sensitive
to a superposed static magnetic field.
\begin{figure}
\centerline{\includegraphics[width=\linewidth,keepaspectratio]{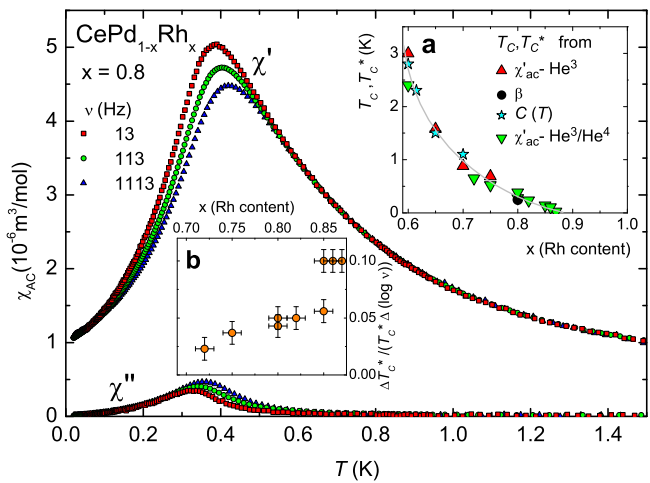}}
\caption{(color online) Ac susceptibility of a single crystal
($x=0.8$) for 3 selected frequencies in a modulation field of
$\mu_{0}h_{ac}=11\,\mu$T. Inset a: Phase diagram of CePd$_{1-x}$Rh$_x$
for $x \geq 0.6$. Inset b: Relative temperature shift of the
maximum in $\chi '(T)$ per frequency decade $\Delta
T_C^*/(T_C^*\Delta(\textrm{log}\, \nu))$ as a function of Rh
content.} \label{fig1}
\end{figure}
In fact, only 15~mT are sufficient to depress the absolute
value to 3/5 of the signal in zero field. However, examining the
relative temperature shift per decade in $\nu$ vs. $x$
(inset b of Fig.~2), we find that this shift of about 3 to 10\%
per decade is considerably larger than in canonical metallic spin
glasses which exhibit only 1 to 2\%. The observed shift is
similar in magnitude to the one observed in insulating spin
glasses, but well below the value of about 28\% observed in a
superparamagnet~\cite{Mydosh}. %In addition, the maximum expected
%in $C(T)$ at a temperature somewhat higher than $T_C^*$, as in spin
%glasses, has been observed for $x=0.8$ at $T\approx T_C^*$ and
%remarkably not seen for $x\geq 0.85$, where $C(T)/T$ diverges
%without any sign of transition~\cite{Pikul06}.

The frequency dependence of $\chi_{ac}(T)$ provides evidence for the
existence of clusters in the system. The change of the magnitude of
the shift suggests that the properties of the clusters, e.\,g.,
their size and/or coupling strength, vary with the Rh content. In
fact, at $x\approx 0.85$ a rapid change of $T_K$ was
observed~\cite{Sereni07}. Very likely, the random distribution of Rh
and Pd ligands creates regions with different local $T_K$ values,
due to  differences in the hybridization of the Cerium 4$f$
electrons with the valence electrons of these differing ligands.
While Pd nearest neighbors tend to stabilize the Ce-moment, Rh
ligands seem to screen it. The strength of the Kondo screening on a
given Ce site thus depends sensitively on the local environment.
This is in agreement with the analysis of $C(T)$~\cite{Pikul06}.
Since the Kondo interaction is rather extended across the lattice,
this effect has to be interpreted in a different way than
percolation effects caused by the dilution of magnetic moments. For
$0.8\leq x\leq0.87$, the dimensionless Sommerfeld-Wilson ratio
$R_{W}=\frac{\chi/\chi_0}{\gamma/\gamma0}$ gives values between 20
and 30, leading to an estimated typical cluster size of about 5
spins ~\cite{Miranda05}.
%
%------magnetization-----------------------------------------------------------------
\begin{figure}
\centerline{\includegraphics[width=\linewidth,keepaspectratio]{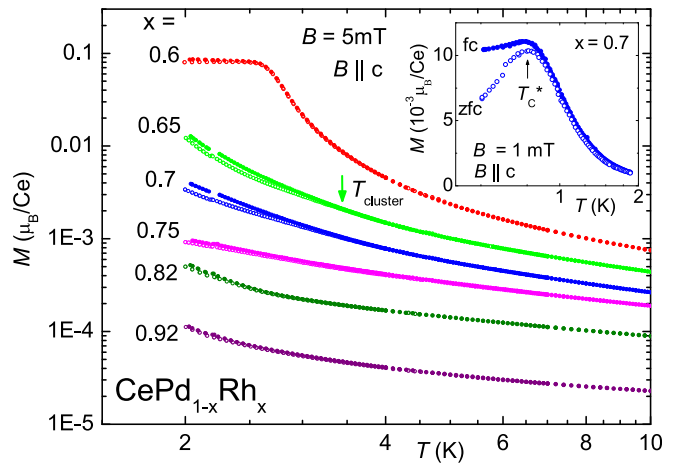}}
\caption{(color online) Dc magnetization $M$ as a function of $T$
in a constant field of 5~mT. FC (thick) and ZFC
(thin) data on single crystals with Rh contents $0.6\leq x\leq
0.92$ are shown down to 2~K. The onset of the temperature
hysteresis, related to the formation of clusters (see text), is
marked by the green arrow (exemplarily shown for $x=0.65$).
There is no cluster freezing for the sample with $x=0.6$ above its $T_{C}$.
Inset: low-$T$ data for $x=0.7$ measured at $\mu_{0}H=1$~mT.} \label{fig2}
\end{figure}
To confirm the existence of freezing clusters, dc magnetization $M$
was measured as a function of temperature on single crystals within
the Rh concentration range $0.6\leq x\leq 0.92$. The inset of Fig.~3
shows the results of field-cooled (FC) and zero-field-cooled (ZFC)
measurements for $x=0.7$ and $\mu_{0}H=1$~mT. Below the freezing
temperature, a clear deviation is observed between FC and ZFC: While
the FC curve saturates below $T_C^*$, the ZFC one exhibits a cusp at
$T_C^*$. This demonstrates the irreversibility of the freezing
process in agreement with our $\chi_{ac}(T)$ results. Remarkably, a
small difference between the FC and ZFC curves exists also at much
higher temperatures (cf. the main part of Fig.~3). We associate the
temperature below which this irreversibility is observed with
$T_{cluster}$, i.\,e., the characteristic temperature for the
formation of short-range order in clusters. With increasing $x$, the
low-$T$ magnetization decreases by several orders of magnitude,
indicative of a drastic reduction of the average moment per Ce-site
and consistent with the strong reduction of the magnetic
entropy~\cite{Pikul06}.
%
%------discussion & scheme------------------------------------------------------------------------
\begin{figure}
\centerline{\includegraphics[width=\linewidth,keepaspectratio]{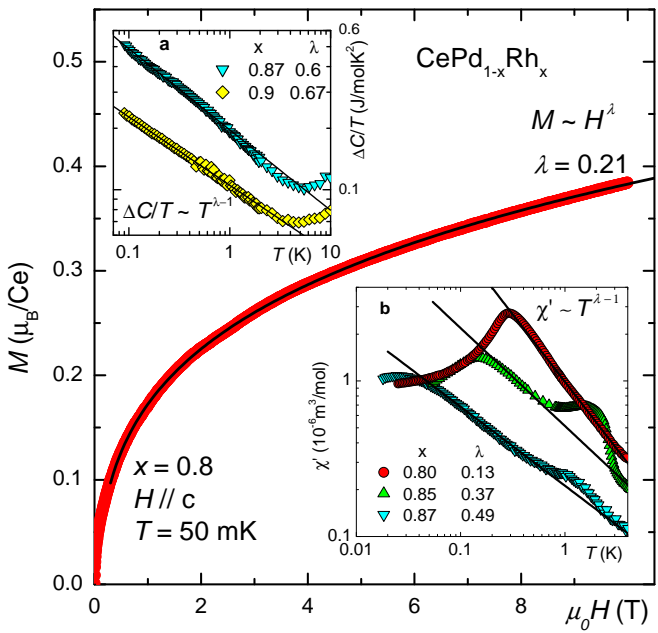}}
\caption{(color online) Field dependence of the magnetization $M$
for a single crystal with $x=0.8$ at 50~mK. The data follow a
power-law function $M\propto H^{\lambda}$ (with $\lambda =0.21$).
Inset a: $4f$ contribution to the specific heat for polycrystals of
$x=0.87$ and $x=0.9$~\cite{Pikul06} as $\Delta C/T$ vs. $T$ on
double-log scale. Solid lines indicate power-law behavior. Inset b:
ac susceptibility $\chi'$ of three polycrystals. The lines indicate
fits with $\chi'\propto T^{\lambda-1}$ above $T_C^*$. The curves for
sample $x=0.85$ and 0.87 present two humps, which are due to
impurity phases.}\label{fig3}
\end{figure}

The following scenario may account for all our findings: At
temperatures high enough to overcome the Kondo screening,
fluctuating magnetic moments exist on every Ce site; below the
average Kondo temperature $\langle T_K\rangle$, an increasing number
of $f$-moments becomes screened; however, due to the statistical
distribution of Rh dopands on the Pd site and the strong dependence
of the local $T_K$ on the number of Pd nearest neighbors, there
remain regions where the Kondo scale has not yet been reached;
inside these regions, the $f$-moments are still unscreened; at even
lower temperatures, $T<T_{cluster}$, these moments form clusters
with predominantly FM coupling of the moments; within this
temperature regime, the clusters are fluctuating independently; on
further cooling below $T_C^*$, random freezing of the cluster
moments sets in, leaving a static spin configuration. Such a
scenario is compatible with all our observations: (i) The formation
of clusters, (ii) their freezing, (iii) the small entropy at low
temperatures and (iv) the negative sign of the thermal expansion,
which points to short-range ordering even at temperatures much above
$T_C^*$. As the broad distribution of local Kondo temperatures is
responsible for the cluster formation, we propose to call the
low-temperature state in CePd$_{1-x}$Rh$_x$ a "Kondo-cluster glass".

The decrease in concentration of the unscreened moments, along with
the small cluster size, might explain why the freezing has been
detected by $\chi_{ac}(T)$ down to very low $T$ (for large $x$), but
was not seen in other techniques, e.g., $\mu$SR or specific heat
\cite{Adroja}. CePd$_{1-x}$Rh$_x$ is different from other Ce-systems
like CeCu$_{1-x}$Ni$_x$~\cite{Marcano07}, where $T_K$ is small in
the entire composition range and the freezing could be observed by
specific heat as well as by $\mu$SR experiments for all
relevant concentrations. %In the latter case a percolative cluster
%scenario has been proposed~\cite{Marcano07}.

%The NFL behavior evident in the specific-heat results on
%CePd$_{1-x}$Rh$_x$ has previously been ascribed to the broad
%distribution of single-ion Kondo temperatures~\cite{Pikul06}. The
%negative sign of $\Gamma$, however, proves that such a scenario is
%insufficient.
Since the magnetic measurements reveal the existence of clusters,
the observed NFL behavior may be described by the quantum Griffiths
phase scenario~\cite{CastroNeto98,CastroNeto00}, which predicts
% the observed weak power-law divergence in $C(T)/T$ in
%CePd$_{1-x}$Rh$_x$ at $x\geq 0.85$ might alternatively be explained
%within the quantum Griffiths phase
%scenario~\cite{CastroNeto98,Dobro05}.
%scenario~\cite{CastroNeto98,Dobro05,Salamon02,Ouyang06,Guo08}.
%Striking evidence for a Griffiths phase, which predicts
$\chi'\propto C/T\propto T^{\lambda - 1}$ and $M\propto
H^{\lambda}$, with $0\leq \lambda \leq 1$. As shown in Fig.\,4, both
the specific heat coefficient $\Delta C(T)/T$ and susceptibility
$\chi'(T)$ follow a power-law behavior well above $T_C^*$, where the
exponent varies systematically with $x$. Moreover, the
field-dependent magnetization for a single crystal with $x=0.8$ at
50~mK (below $T_C^*$) follows a power-law function with $\lambda
=0.21$, satisfactorily close to that found in $\chi'(T)$. A tiny
hysteresis can be also observed with a coercive field of about 5~mT,
but no step-like behavior can be resolved, in contrast to what has
been seen in CeCu$_{1-x}$Ni$_x$~\cite{Marcano07}.

To conclude, the lack of a divergence of the Gr\"uneisen ratio
excludes a standard QCP in CePd$_{1-x}$Rh$_x$ and raises question
about the origin of the pronounced NFL behavior. Whereas weak
power-law divergences in the specific-heat coefficient may be
considered as being due to a single-ion effect originating in the
broad distribution of local Kondo temperatures~\cite{Pikul06}, the
observed negative sign of the thermal expansion strongly points to a
cooperative effect. The detailed investigation of magnetic
properties close to the disappearance of magnetic order reveals the
formation of a "Kondo-cluster-glass" state, where the clusters
result from regions of low local Kondo temperatures~\cite{Dobro05}.
NFL effects in the specific heat, susceptibility and magnetization
have been found to be compatible with the quantum Griffiths phase
scenario.

We are grateful to J. G. Sereni, Q. Si and T. Vojta for helpful
conversations. A. P. acknowledges a fellowship by the Alexander
von Humboldt Foundation. This work was supported by the DFG
Research Unit 960 "Quantum Phase Transitions".


\begin{thebibliography}{40}
\bibitem{Loehneysen07} H. v. L\"ohneysen {\it et al.}, Rev. Mod. Phys. {\bf 79}, 1015 (2007).
\bibitem{Stewart} G. R. Stewart, Rev. Mod. Phys. {\bf 73}, 797 (2001); {\it ibid.} {\bf 78}, 743 (2006).
\bibitem{Sullow} S. S\"{u}llow, M.C. Aronson, B.D. Rainford, P. Haen, Phys. Rev. Lett. {\bf 82}, 2963 (1999).
\bibitem{Eichler} T. Burghardt, H. Neemann, E. Bauer, A. Eichler, J.
Phys.: Condes. Matter {\bf 17}, 871 (2005).
\bibitem{Larrea} J. Larrea {\it et al.}, Phys. Rev. B {\bf 72}, 035129 (2005).
\bibitem{Pfleiderer97} C. Pfleiderer {\it et al.}, Phys. Rev. B {\bf 55}, 8330 (1997).
\bibitem{Saxena00} S. S. Saxena {\it et al.}, Nature {\bf 406}, 587 (2000).
\bibitem{Uhlarz04} M. Uhlarz {\it et al.}, Phys. Rev. Lett. {\bf 93}, 256404 (2004).
\bibitem{Kirkpatrick03} T. R. Kirkpatrick and D. Belitz, Phys. Rev. B {\bf 67}, 024419 (2003).
\bibitem{Chubukov04} A. V. Chubukov, C. Pepin and J. Rech, Phys. Rev. Lett. {\bf 92}, 147003 (2004).
\bibitem{Gegenwart08} P. Gegenwart, Q. Si and F. Steglich, Nature Physics {\bf 4}, 186 (2008).
\bibitem{Vojta03} T. Vojta, Phys. Rev. Lett. {\bf 90}, 107202 (2003).
\bibitem{Sokolov06} D. A. Sokolov {\it et al.}, Phys. Rev. Lett. {\bf 96}, 116404 (2006).
\bibitem{Huy07} N. T. Huy {\it et al.}, Phys. Rev. B {\bf 75}, 212405 (2007).
\bibitem{Kappler91} J. P. Kappler {\it et al.}, Physica B {\bf 171}, 346 (1991).
\bibitem{Sereni07} J. G. Sereni {\it et al.}, Phys. Rev. B {\bf 75}, 024432 (2007).
\bibitem{Pikul06} A. P. Pikul {\it et al.}, J. Phys. Condens. Matter {\bf 18}, L535 (2006).
\bibitem{Deppe06} M. Deppe {\it et al.}, Physica B {\bf 378-380}, 96 (2006).
\bibitem{Zhu} L. J. Zhu {\it et al.}, Phys. Rev. Lett. {\bf 91}, 066404 (2003).
%\bibitem{Command} This holds true for any pressure-sensitive QCP [5]. To our knowledge, every HF QCP established so far, belongs to this category.
%\bibitem{Kuechler} R. K\"uchler, Dissertation, TU Dresden (2005), unpublished.
\bibitem{Mydosh} J. A. Mydosh, {\it Spin Glasses - An Experimental Introduction}, Taylor and Francis (1993).
\bibitem{Miranda05} E. Miranda and V. Dobrosavljevi\'{c}, Rep. Prog. Phys. {\bf 68}, 2337 (2005).
\bibitem{Adroja} D. T. Adroja {\it et al.}, Phys. Rev. B {\bf 78}, 014412 (2008).
\bibitem{Marcano07} N. Marcano {\it et al.}, Phys. Rev. Lett. {\bf 98}, 166406 (2007).
\bibitem{CastroNeto98} A. H. Castro Neto, G. Castilla, and B. A. Jones, Phys. Rev. Lett. {\bf 81}, 3531 (1998).
\bibitem{CastroNeto00} A. H. Castro Neto and B. A. Jones, Phys. Rev. B {\bf 62}, 14975 (2000).
\bibitem{Dobro05} V. Dobrosavljevi\'{c} and E. Miranda, Phys. Rev. Lett. {\bf 94}, 187203 (2005).
%\bibitem{Salamon02} M. B. Salamon, P.Lin, and S. H. Chun, Phys. Rev. Lett. {\bf 88}, 197203 (2002).
%\bibitem{Ouyang06} Z. W. Ouyang {\it et al.}, Phys. Rev. B {\bf 74}, 094404 (2006).
%\bibitem{Guo08} S. Guo {\it et al.}, Phys. Rev. Lett. {\bf 100}, 017209 (2008).


\end{thebibliography}
\end{document}